\input{aipcheck}

\documentclass[
    ,final            % use final for the camera ready runs
  ]
  {aipproc}

\layoutstyle{6x9}

\begin{document}

\title{Spectral evidence for jets from Accreting Millisecond X-ray Pulsars}


\author{David M. Russell}{
  address={Astronomical Institute `Anton Pannekoek', University of Amsterdam, Kruislaan 403, 1098 SJ Amsterdam, the Netherlands},
  email={D.M.Russell@uva.nl}
}

\author{Rob P. Fender}{
  address={School of Physics \& Astronomy, University of Southampton, Southampton, SO17 1BJ, UK}
}

\author{Peter Jonker}{
  address={SRON, Netherlands Institute for Space Research, Sorbonnelaan 2, 3584 CA Utrecht, Netherlands}
  ,altaddress={Harvard-Smithsonian Center for Astrophysics, 60 Garden Street, Cambridge, MA 02138, USA} % additional visiting address  
}

\author{Dipankar Maitra}{
  address={Astronomical Institute `Anton Pannekoek', University of Amsterdam, Kruislaan 403, 1098 SJ Amsterdam, the Netherlands}
}

\begin{abstract}

Transient radio emission from X-ray binaries is associated with synchrotron emission from collimated jets that escape the system, and accreting millisecond X-ray pulsars (AMXPs) are no exception.  Although jets from black hole X-ray binaries are well-studied, those from neutron star systems appear much fainter, for reasons yet uncertain.  Jets are usually undetectable at higher frequencies because of the relative brightness of other components such as the accretion disc.  AMXPs generally have small orbital separations compared with other X-ray binaries and as such their discs are relatively faint.  Here, I present data that imply jets in fact dominate the radio-to-optical spectrum of outbursting AMXPs.  They therefore may provide the best opportunity to study the behaviour of jets produced by accreting neutron stars, and compare them to those produced by black hole systems.

\end{abstract}

\maketitle

%%%%%%%%%%%%%%%%%%%%%%%%%%%%%%%%%%%%%%%%%%%%
%% MAINMATTER
%%%%%%%%%%%%%%%%%%%%%%%%%%%%%%%%%%%%%%%%%%%%

\section{Introduction}

Although accreting millisecond X-ray pulsars (AMXPs) are inevitably most studied at X-ray energies, an additional wealth of information regarding the nature of these systems is manifested in their lower energy radiation. In the optical domain, thermal emission from the outer regions of the accretion disc (from the reprocessing of illuminating X-ray photons) is seen during outbursts, much like other low-mass X-ray binaries (LMXBs) \cite{wanget01}. At still lower energies, transient radio emission has been detected from AMXPs during outburst which, like other LMXBs, likely originates in synchrotron emission from collimated, compact jets that escape the binary \cite{miglfe06}. Jets formed by accretion onto neutron stars (NSs) are relatively poorly understood compared with those of accreting black holes (BHs). This is likely because the radio emission from NS jets appears to be intrinsically fainter than that produced by stellar-mass BHs, and accreting NSs with the strongest magnetic fields have no radio emission associated with jets \cite{miglfe06,masska08}. Nevertheless, relativistic jets have been directly resolved in a few NS XBs \cite{fomaet01,tudoet08}.

The existence of jets produced by accreting NSs, and the amount of accretion energy they channel back into the interstellar medium, are important for understanding the process of accretion onto these objects. In BH XBs, the jet power can be comparable to the bolometric X-ray luminosity \cite{gallet05,russet07cyg} but in the NS systems the power contained in the jets may be much lower, but is currently uncertain. Accreting, low-magnetic field NS XBs can be separated into three classes \cite{vand06} according to their X-ray properties: Z-sources, atolls and AMXPs. Transient radio emission has been associated with all three classes \cite{miglfe06}. For AMXPs, jets are important because they provide a source of angular momentum loss of the accreting matter, which may significantly influence the evolution of the NS spin.

%There are two main methods to constrain the accretion power contained in NS jets. The first is to study how the jets influence the surrounding matter \cite{tudoet06} and the second is to use the luminosity of the compact jet. In practice, the first method is not possible for most LMXBs because they reside in low density environments \cite{hein02}. The second method is possible from the core radio luminosity if a radiative efficiency for the emission is assumed. In addition to the efficiency, the broadband spectrum of the jet must be known (either assumed, or observed) because the total power is dominated by the higher energy emission. This is true at least if the spectrum produced by the jet is optically thick at low frequencies, resulting in an approximately flat ($\alpha = 0$, where $F_{\nu} \propto \nu^{\alpha}$) spectrum that breaks to an optically thin spectrum (with $\alpha < 0$) at some higher frequency. This break frequency resides close to the near-infrared (NIR) regime for BH X-ray binaries \cite{corbfe02,homaet05,russet06} but may be further into the mid-infrared (MIR) in NS systems \cite{miglet06}.

One method to constrain the accretion power contained in NS jets is to use the luminosity of the compact jet, if a radiative efficiency for the emission is assumed. In addition to the efficiency, the broadband spectrum of the jet must be known (either assumed, or observed) because the total power is dominated by the higher energy emission. This is true at least if the spectrum produced by the jet is optically thick at low frequencies, resulting in an approximately flat ($\alpha = 0$, where $F_{\nu} \propto \nu^{\alpha}$) spectrum that breaks to one which is optically thin (with $\alpha < 0$) at some higher frequency. This break frequency resides close to the near-infrared (NIR) regime for BH XBs \cite{corbfe02,homaet05,russet06} but may be further into the mid-infrared (MIR) in NS systems \cite{miglet06}. In these proceedings we compile radio, NIR and optical data of AMXPs and constrain the dominating emission processes from their spectral energy distributions (SEDs). These are compared to data in the literature of other types of accreting NS as well as BH systems.

%%%%%%%%%%%%%%%%%%%%%%%%%%%%%%%%%%%%%%%%%%%%
%% Sample figure:
%%
%% The option [height=...] scales the picture to the given height,
%% without it it would be printed at its nominal size
%%%%%%%%%%%%%%%%%%%%%%%%%%%%%%%%%%%%%%%%%%%%

\begin{figure}
  \includegraphics[height=.45\textheight,angle=270]{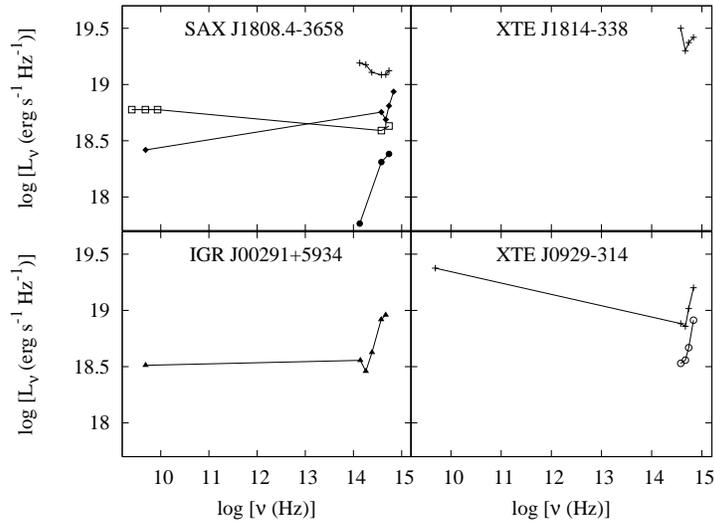}
  \caption{Radio-to-optical spectral energy distributions of four AMXPs.}
\end{figure}

\section{Data collection and analysis}

The data for this study were collected from the literature; most were previously compiled for two published works \cite{russet06,russet07NSs}; here additional radio data of AMXPs (from refs \cite{gaenet99,rupeet02,rupeet04,rupeet05}) were obtained to construct radio-to-optical SEDs. Optical and NIR magnitudes and fluxes were de-reddened using the estimated values of the interstellar extinction towards each source and all fluxes were converted to monochromatic luminosities (flux density scaled with distance; this is used to compare between sources) adopting their estimated distances from the literature (see Table 1 of \cite{russet07NSs} for distance and extinction estimates).

The SEDs of four AMXPs are presented in Fig. 1. The data for individual SEDs were taken within two days; i.e. they are quasi-simultaneous. In the optical region, the spectrum of the X-ray heated accretion disc (with spectral index $\alpha > 0$) can be seen (see e.g. \cite{wanget01} for discussions) in all four AMXPs. The SEDs of all four systems are also consistent with there being a NIR-excess above the disc emission at high luminosities, which disappears at low luminosities. A radio source is detected at high luminosities when the NIR-excess is present. The radio--optical spectral index is close to zero (ranging between $\alpha$ = -0.10 and +0.07). This implies that either the jet break (from optically thick to optically thin emission) is close to the NIR regime, or the break is further into the MIR and the optically thick spectrum is itself more inverted ($\alpha$ > 0).

Further evidence for the transient nature of the NIR-excess is shown in Fig. 2. Here, an optical colour-magnitude diagram (CMD) of the 2002 outburst of XTE J0929--314 is presented. During an outburst, the accretion disc is hotter, and therefore bluer, at high luminosities. We overplot a model for a simple, single-temperature heated black body (see \cite{maitba08}). The model is able to reproduce the overall relation between colour and magnitude, except at the highest flux levels, in which the data are redder than expected. From the SEDs, which show a NIR-excess from the jet at the highest luminosities, it is clear that optical--NIR CMDs of outbursts can be successfully used to separate the disc and jet components.

\begin{figure}
  \includegraphics[height=.34\textheight,angle=0]{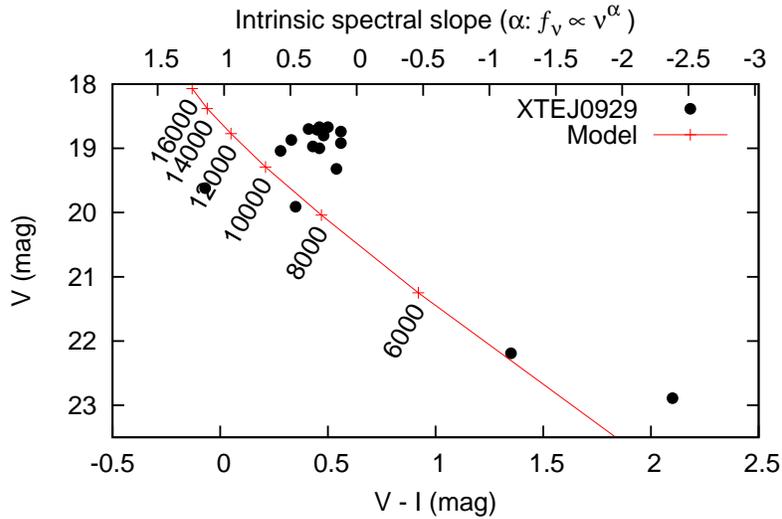}
  \caption{CMD of XTE J0929--314 from its 2002 outburst and quiescence (data are from \cite{gileet05}).}
\end{figure}

\section{Discussion and Conclusions}

By comparing these results with similar analyses from the literature of other NS XBs, it is clear that AMXPs most likely have the most `optically visible' jets -- the jet / disc ratio in the optical regime is larger in AMXPs than for other accreting NS XBs. There is little evidence for jets contributing to the optical light of Z-sources (which have long orbital periods and therefore larger accretion discs) \cite{russet07NSs} even though they are more luminous in radio than AMXPs \cite{miglfe06}. A recent SED shows the jet and disc emission may be of equal flux density in the MIR for one Z-source, Sco X--1 \cite{vrti08} whereas for the atoll 4U 0614+09 this lies in the NIR \cite{miglet06}. Optical--NIR CMDs of eight outbursts of the atoll Aql X--1 (which has a long orbital period) show no evidence for a jet contribution \cite{maitba08}. AMXPs have in general the shortest orbital periods and hence smaller discs, so it may be that this is the reason for AMXPs to have the highest jet / disc ratios in the optical regime (some of the strongest evidence for optical jets in BH systems comes from XTE J1118+480 which also has a short orbital period). Alternatively, the difference may originate in a different break frequency in the jet spectrum between AMXPs, Z-sources and atolls. Whatever the reason, it appears that the best examples for studying jets in accreting NSs at frequencies higher than radio, are AMXPs.

%%%%%%%%%%%%%%%%%%%%%%%%%%%%%%%%%%%%%%%%%%%%%%%%
%% BACKMATTER
%%%%%%%%%%%%%%%%%%%%%%%%%%%%%%%%%%%%%%%%%%%%%%%%

%\begin{theacknowledgments}

%\end{theacknowledgments}



\IfFileExists{\jobname.bbl}{}
 {\typeout{}
  \typeout{******************************************}
  \typeout{** Please run "bibtex \jobname" to optain}
  \typeout{** the bibliography and then re-run LaTeX}
  \typeout{** twice to fix the references!}
  \typeout{******************************************}
  \typeout{}
 }

\end{document}